# Graph neural network coarse-grain force field for the molecular crystal RDX

Brian H. Lee,[1] James P. Larentzos,[2] John K. Brennan,[2] Alejandro Strachan[1, a]

Author affiliations
[1]*School of Materials Engineering and Birck Nanotechnology Center, Purdue University, West Lafayette, Indiana 47907, USA*
[2]*U.S. Army Combat Capabilities Development Command (DEVCOM) Army Research Laboratory, Aberdeen Proving Ground, Maryland 21005, USA*

Author email
[a]Author to whom correspondence should be addressed: strachan@purdue.edu

## Abstract

Condense phase molecular systems organize in wide range of distinct molecular configurations, including amorphous melt and glass as well as crystals often exhibiting polymorphism, that originate from their intricate intra- and intermolecular forces. While accurate coarse-grain (CG) models for these materials are critical to understand phenomena beyond the reach of all-atom simulations, current models cannot capture the diversity of molecular structures. We introduce a generally applicable approach to develop CG force fields for molecular crystals combining graph neural networks (GNN) and data from an all-atom simulations and apply it to the high-energy density material RDX. We address the challenge of expanding the training data with relevant configurations via an iterative procedure that performs CG molecular dynamics of processes of interest and reconstructs the atomistic configurations using a pre-trained neural network decoder. The multi-site CG model uses a GNN architecture constructed to satisfy translational invariance and rotational covariance for forces. The resulting model captures both crystalline and amorphous states for a wide range of temperatures and densities.

# 1. Introduction

Molecular crystals are an important class of materials widely used in the pharmaceutical[1] and electronics industries,[2–4] as well as high-energy-density materials.[5–7] Molecular modeling plays a key role in the development of a fundamental understanding of these materials and the discovery of new ones with desirable properties.[8–12] While all-atom molecular dynamics (MD) enables an accurate description of the thermo-mechanical and chemical properties of molecular materials, it remains restricted to sub-micron and sub-millisecond scales, limiting the phenomena that can be described. Thus, particle-based, coarse-grain (CG) modeling is central to connecting the microscopic world to mesoscales and atomistics to the continuum.[13–18]

A critical target of CG models is accurate description of the molecular materials' structural arrangements that can span from crystalline to amorphous configurations depending on the thermodynamic conditions. Accurate CG descriptions of amorphous polymers have contributed significantly to our understanding of glassy physics[19,20] and biomolecules.[15,21,22] Unfortunately, less progress has occurred in CG representations of molecular crystals, where many molecules of interest form open, low-symmetry crystal structures. Most of the available CG models, with the exception of a CG model for water,[23] cannot capture both crystal and melt phases and the transformation between them.

For example, multiple CG models[17,24–28] have been developed for the high-energy-density material 1,3,5-Trinitro-1,3,5-triazinane (RDX) using the multiscale coarse-graining (MSCG/FM) approach, which utilized spline functions to match the forces between groups of atoms.[13,15,17,25,29] These models can capture either the crystalline or the molten state depending on the dataset used to develop them, but for a given model have been unable to describe the structural characteristics of both states correctly. Specifically, for the so-called RDX-T model[30], developed by matching the atomistic forces of a mainly liquid phase dataset of RDX, the mechanical properties as well as the structural properties of the amorphous state of RDX can be correctly captured. However, the crystal state for the RDX-T model is captured as an affinely deformed hcp structure rather than the *Pbca* structure of the reference atomistic model. Moreover, the melt transition is overestimated for the RDX-T model. A more recently developed model, the so-called RDX-TCDD model,[25] accurately attain the *Pbca* crystal structure and exhibit plastic deformation that is equivalent to the reference atomistic model. However, the liquid phase is more structured with peaks in the radial distribution functions matching those observed from the α-RDX crystals.

In this work, we propose a coarse grain model for RDX based on a graph neural network (GNN) that captures both the low-symmetry molecular crystal state as well as amorphous configurations of RDX. GNNs provide a natural representation of molecular structures, with nodes representing atoms (or groups of atoms in CG descriptions) and edges representing interactions and used to share information about the local environment around atoms.[31,32] Recent work has shown the significant power of GNNs to represent atomistic force fields[33–38] trained from extensive sets of accurate electronic structure calculations. However, GNNs are only emerging in the field of CG models[39–41] with no applications to molecular crystals to date. One of the main challenges in the development of CG GNN models, or any molecular model based on machine learning, is the generation of appropriate training sets. The development of machine learning interatomic potentials (ML-FFs) often involves an iterative approach, where configurations explored by molecular dynamics simulations with ML-FFs are added to the dataset to train new generations of ML-FFs. This process is important because it *teaches* the model about improbable and high-energy configurations that can lead to catastrophic failures if they are not included in the training dataset.

Unfortunately, the approach does not translate directly to CG models, as the reconstruction of an atomistic model from a CG model trajectory is not trivial. We address this challenge by developing a neural network decoder that reconstructs atomistic configurations from their CG representations, allowing expansion of the training dataset for iterative training workflow.

In addition, we formulated two custom GNN architectures to describe inter- and intra-molecular interactions of RDX with forces that are invariant to rigid translations and covariant to rigid rotations, which guarantees the conservation of total linear and angular momenta. We trained the model, named CG-GNNFF, on approximately $10^6$ atomistic configurations. CG-GNNFF correctly reproduces the RDX structure for both crystalline and amorphous states at a wide range of temperatures and densities. In addition, the versatility of the model is demonstrated by extrapolating the model to simulations of faceted nanoparticles that were not included in the training dataset.

## 2. Results

*2.1. Coarse grain graph neural network force field model*

The CG-GNNFF architecture is shown in Fig. 1. For the RDX molecule, the atomistic configuration is coarse-grained into a 4-site model using the center of masses of the three nitro groups and the triazine ring, as depicted in Fig. 1a. The CG interactions are represented by two distinct graphs ($\mathcal{G}_{\text{intra}}, \mathcal{G}_{\text{inter}}$) and the associated GNNs to describe forces arising from the intra- and inter-molecular interactions of the atoms ($F_{\text{intra}}, F_{\text{inter}}$), respectively. This choice, similar to the approach taken by Ruza et al.,[41] was made because the atomistic forces for the intra- and inter-molecular interactions deviate significantly from each other in terms of both the connectivity and the underlying force field.

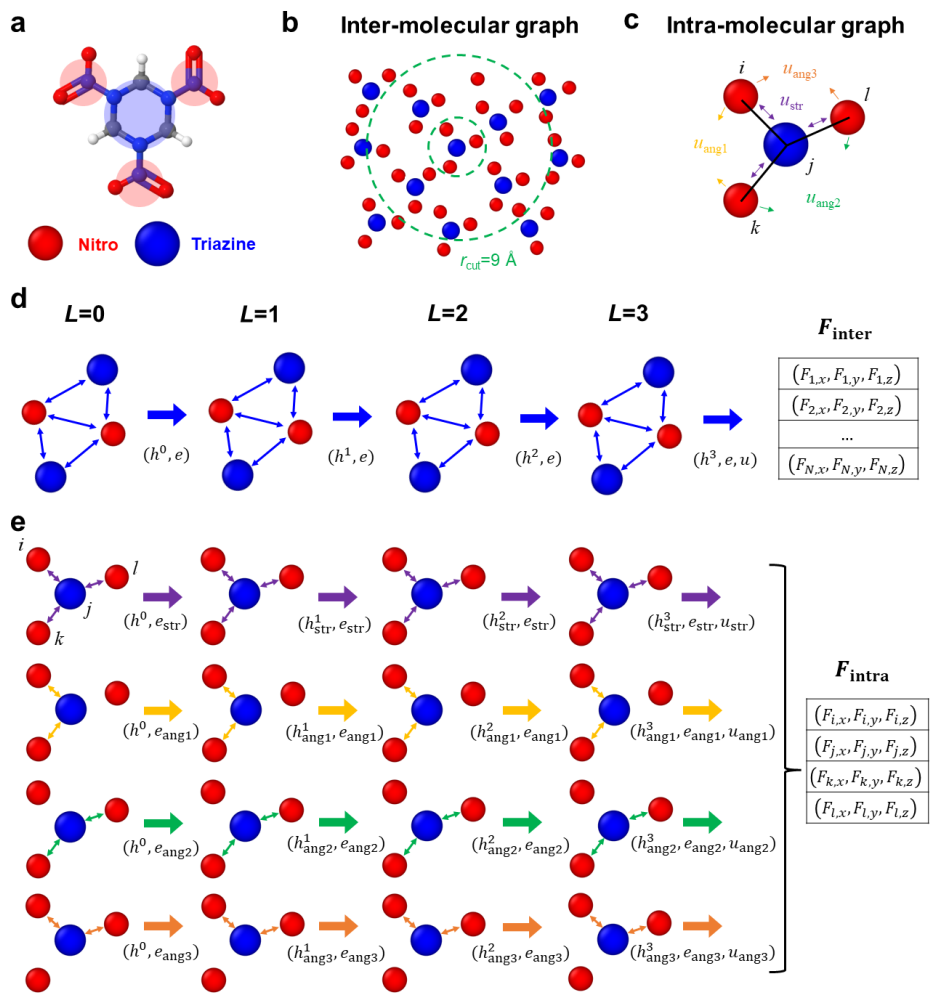

**Figure 1**. Schematics of the CG-GNNFF. (a) An RDX molecule is coarse grained into a 4-site model with 3 nitro and 1 triazine groups. (b) Inter-molecular graphs are made with connectivity specified between all CG particles within cutoff distances, excluding the beads that are within the same molecule. (c) Intra-molecular graphs specify connectivity between nitro and triazine groups of an RDX molecule. (d, e) Message-passing process for (d) inter- and (e) intra-molecular graphs. Node feature, edge feature, direction vector, GNN layer are denoted as $h$, $e$, $u$, and $L$, respectively. $\boldsymbol{F}_{inter}$ and $\boldsymbol{F}_{intra}$ are the output of the GNN.

$\mathcal{G}_{\text{inter}}(\mathcal{V}_{\text{inter}}, \mathcal{E}_{\text{inter}}, \mathcal{U}_{\text{inter}})$, illustrated in Fig. 1b, describes inter-molecular interactions via many-body interactions along each pair of particles within a cutoff distance. The nodes $v_{i,\text{inter}} \in \mathcal{V}_{\text{inter}}$ represent each CG bead and are assigned initial node features ($h^0$) corresponding to one-hot vector of two dimensions that indicates whether the node is a nitro or triazine group. The adjacency matrix for $\mathcal{G}_{\text{inter}}$ was constructed by assigning connectivity between all CG beads within a cutoff distance, excluding the beads within the same molecule. We choose 9 Å as the cutoff to include the first three peaks of the radial distribution function of α-RDX, the ambient density molecular crystal configuration of RDX. We also note that the particles receive information beyond the cutoff distance through the multiple layers of message-passing processes in the GNN model, and the CG forces include many-body interactions.

The edge features, $e_{ij,\text{inter}} \in \mathcal{E}_{\text{inter}}$, were determined as a four-dimensional vector that is a combination of Gaussian basis functions of $d_{ij}$, the distance between particle $i$ and $j$. Detailed descriptions of the edge features and Gaussian bases functions are given in Section 1 of the Supplementary Materials (SM).

Lastly, we store a three-dimensional force direction vector ($u_{\text{inter}}$) for each edge, which assigns the direction of force between particle $i$ and $j$, as explained in detail in the Methods section 4.1. The magnitude of this vector is equal to $f_c(d_{ij})$, where $f_c$ is a cutoff function that smoothly decays to 0 at $r_{\text{cut}}$, as defined by Behler et al.[42],

$$f_c(d_{ij}) = \begin{cases} 0.5[\cos(\pi d_{ij}/r_{\text{cut}}) + 1] & \text{for } d_{ij} \leq r_{\text{cut}} \\ 0 & \text{for } d_{ij} > r_{\text{cut}} \end{cases} \quad (1)$$

Such choice ensures that the force smoothly decays to 0 at $r_{\text{cut}}$. Using these input features, we implement four layers of GNN to output the forces on each bead (denoted as L=0, 1, 2, and 3 in Fig. 1d).

The intra-molecular interaction graph $\mathcal{G}_{\text{intra}}(\mathcal{V}_{\text{intra}}, \mathcal{E}_{\text{intra}}, \mathcal{U}_{\text{intra}})$ describes covalent interactions between CG particles in terms of many-body forces along nitro and triazine centers, and angular terms centered around the triazine. The node features are identical to those in $\mathcal{V}_{\text{inter}}$. The connectivity between the nodes is assigned according to the intramolecular connectivity. We describe stretching, angular, and torsional interactions of the underlying bonded forces of the atoms as a combination of four forces (Fig. 1e). The stretching force between nitro and triazine pairs, denoted as $F_{\text{str}}$, are treated by a single graph. The edge features incorporate the distance between nitro-triazine pairs as well as angles associated with nitro-triazine-nitro triplets, as explained in SM section 1. The force direction vector $(u_{\text{str}})$ is defined as $\text{norm}(\boldsymbol{d}_{N-T})$, the normalized displacement vector between the nitro-triazine pair.

The angular interaction forces between the three nitro-triazine-nitro triplets $(F_{\text{ang1}}, F_{\text{ang2}}, F_{\text{ang3}})$ are treated with separate graphs for each triplet. This choice is made to ensure that the node-level GNN output on the nitro groups for each angular interaction are numerically equal. In this case, we can formulate the total torque on the molecule from the intramolecular interactions to conserve the angular momentum with a judicious choice of the force direction vector $(u_{\text{ang}})$. Specifically, we define $u_{\text{ang}}$ as $\text{norm}(\boldsymbol{d}_{N_i-T} \times \boldsymbol{d}_{T-N_j})/|\boldsymbol{d}_{N_i-T}|$ for the edge between nitro particle $i$ and triazine in the nitro $i$−triazine−nitro $j$ triplet. The edge features are explained in detail in SM section 1. The final intramolecular force output is equal to the sum of $F_{\text{str}}$, $F_{\text{ang1}}$, $F_{\text{ang2}}$, and $F_{\text{ang3}}$. Lastly, the intramolecular forces on the triazine group are not directly obtained from the GNN, but computed as the negative sum of the forces on the three nitro groups to ensure conservation of linear momentum.

Using such architecture, we trained the CG-GNNFF models that make node-level predictions of the CG forces by updating the initial node features $(h^0)$ through message passing processes. For the message passing step, we use the graph transformer model[43] described in detail in the Methods section 4.1.

*2.2. Iterative training procedure*

As the ML-FFs are agnostic to the underlying physics of the system except for the symmetries incorporated in the representation of molecular structures, their performances are highly dependent on the datasets used to train the models. To obtain datasets representative of physically relevant configurations of molecules, we utilized the iterative training procedure depicted in Fig. 2. For the initial training dataset, we performed canonical ensemble (NVT) atomistic simulations of initially crystalline molecules at ambient densities and temperatures of 250, 500, and 750 K for 100 ps (Fig. 2a). From these atomistic simulations, we extracted 100 configurations at 1 ps intervals and obtained the mean force acting on each of the four CG beads. This is accomplished via constrained NVT simulations, where the centers of mass of the nitro and triazine groups are fixed at their initial positions (Fig. 2b). The mean atomistic forces $\langle f \rangle$ for the training datasets are obtained from 100 snapshots for 10 ps of constrained NVT simulations. More details of the atomistic simulations are given in the Methods section 4.2.

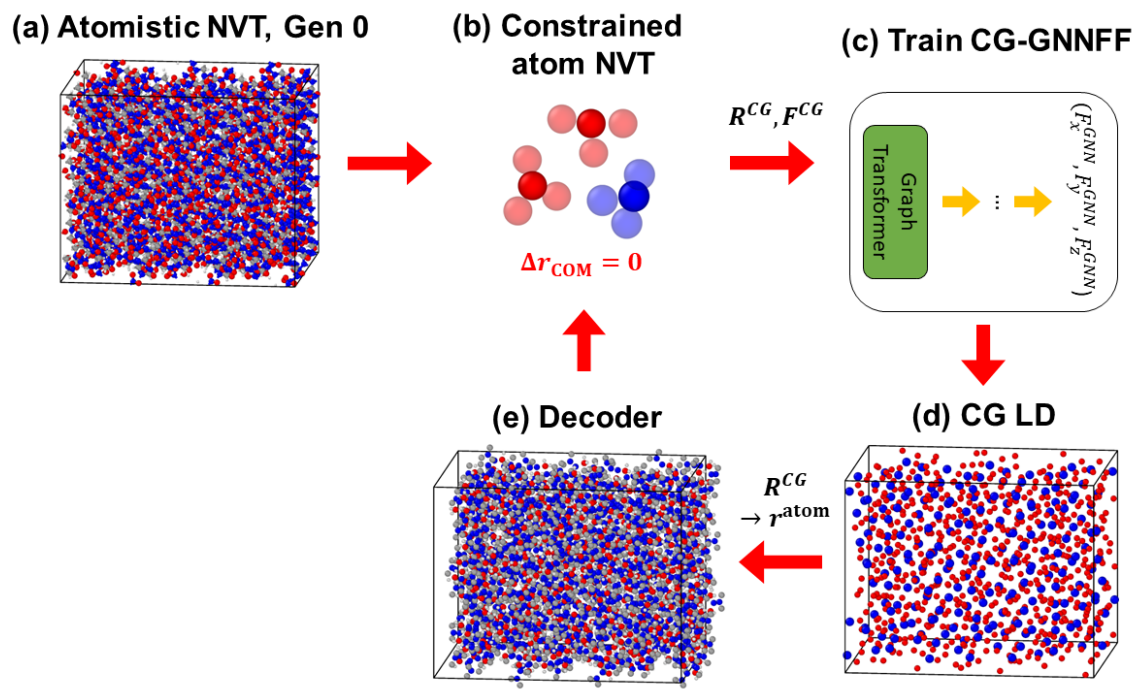

**Figure 2**. Iterative training process. (a) Atomistic NVT simulations are performed for Gen 0. (b) Constrained atomistic simulations are performed in which the center of mass positions of the nitro and triazine groups are fixed in space. (c) CG model positions and forces from constrained atomistic simulations are used to train CG-GNNFF. (d) CG MD simulations with the trained CG-GNNFF are performed. (e) Decoder is used to reconstruct CG model trajectories into atomistic

trajectories. Constrained atomistic simulations on these trajectories are performed to create the next generation of data.

Using the initial training dataset, denoted generation (Gen) 0, we trained a CG-GNNFF (Fig. 2c) and performed a coarse-grain MD simulation with a Langevin thermostat (Fig. 2d), as described in Methods section 4.2. During the dynamical simulation, the CG model explored configurations not included within the initial training data, whereby accurate extrapolation of forces from the GNN model for these configurations is not guaranteed. Typically, for atomistic ML-FFs [44,45] such configurations are added to the training dataset to enhance the models. However, for the CG models, such expansion of the dataset is not trivial as there is no one-to-one correspondence between the CG and atomistic trajectories. While mapping an atomistic to a CG configuration can be computed simply by using the center-of-mass of the group of atoms, the reverse process is a one-to-many problem that does not have an analytical solution.

To overcome this challenge, we use a neural network decoder to reconstruct atomistic configurations from the CG model trajectories (Fig. 2e). The decoder is a deep feed-forward neural network whose inputs are a 12-dimensional vector corresponding to the cartesian coordinates of the 4 CG particles of RDX, and whose outputs are a 63-dimensional vector corresponding to the cartesian coordinates of the 21 RDX atoms. The neural network consists of two hidden layers with 12 and 36 nodes in each layer with leaky ReLu activation functions.[46] The reconstructed atomic positions were adjusted to ensure that the center of masses of the nitro and triazine groups are equal to the CG particle positions. The decoder model was trained on atomistic configurations of the Gen 0 training data. We do not optimize the decoder beyond the Gen 0 dataset, as the current model is sufficient to provide the initial atomistic configurations that are equilibrated through the constrained NVT simulations.

Using the iterative training procedure depicted in Figure 2, we obtained three generations of datasets with 1.6 million data points. A detailed explanation of the datasets is given in SM section 2.

*2.3 Force prediction accuracy*

The intra- and intermolecular GNN models were trained for 3 generations as described in Section 2 of the SM. The model performance for each generation is depicted in Figs. S3 and S4. The force predictions demonstrate significant improvement in accuracy as the model is trained iteratively for Gen 0 and Gen 1. For example, the models trained on Gen 0 data cannot predict the forces exceeding 20 kcal/mol/Å present in Gen 1 data, but the models trained on subsequent generations can accurately capture them. However, training on additional generations after Gen 1 led to small changes in the prediction accuracy of the model. A more detailed description of the generation-dependent force prediction accuracy is given in Section 4 of the SM. The accuracies of the force prediction by the final generation are plotted in Fig. 3. The model performs with equivalent accuracy on both training and test sets, implying that the GNN model captures the underlying relationship between the CG configurations and the forces without overfitting. The force inference displays reasonable accuracies for both inter- (Fig. 3a, b) and intra-molecular (Fig. 3c, d) interactions, with mean absolute errors (MAE) of ~ 5.3 kcal/(mol·Å) and ~ 3.1 kcal/(mol·Å), respectively. While a direct comparison with other CG RDX models[17,24] is not possible as the force prediction accuracies of these models are not reported, the errors are significantly smaller than the values for other CG ML-FFs[41,47] that reported ~360 (kcal/mol/Å)$^2$ of mean squared errors.

We also note that the theoretical minimum error of a CG model is limited by the noise in the data as the CG model forces represent the mean force ⟨**f**⟩ from a trajectory of atomistic forces as explained by Wang et al.[47] This contrasts with the atomistic force fields trained on quantum mechanical data that lead to one-to-one correspondence in the atomic configurations and their energies. As each CG model configuration can represent multiple atomistic configurations, there is inherent uncertainty in the target CG model forces due to finite sampling, whereby even in ideal cases the CG models cannot attain zero error. Specifically, the average standard deviations of the intra- and inter-molecular forces of the Gen 0 data were 18.43 and 4.76 kcal/(mol·Å), respectively, over the 100 measurements taken over 10 ps intervals. Using a confidence interval of 95%, the margins of error correspond to 3.6 and 0.9 kcal/(mol·Å), respectively, for intra- and inter-molecular forces. Such noise in the ground-truth data limits the prediction accuracy of the CG-GNNFF.

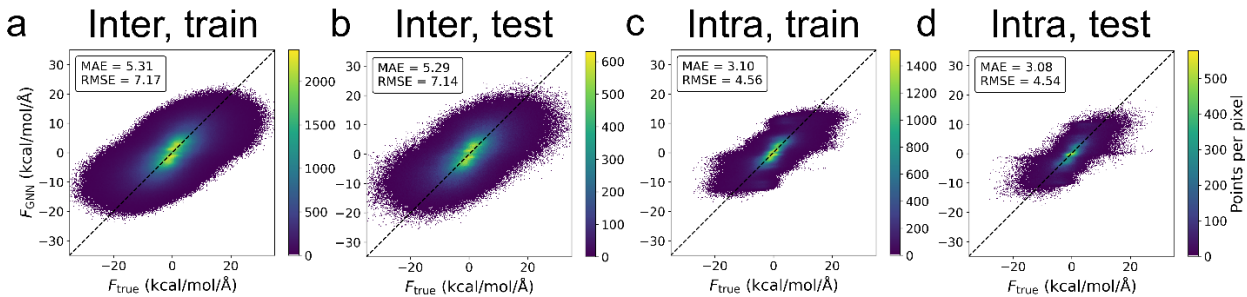

**Figure 3**. Force prediction accuracy for intermolecular training (a), testing (b) and intramolecular training (c) and testing (d) datasets.

*2.4 Structural properties for bulk simulations*

Reproducing the correct structural properties over a wide range of state conditions is a challenging task for CG force fields.[48] To evaluate the ability of CG-GNNFF to describe the structure of RDX, we first compare the crystalline structures obtained from CG and atomistic simulations at room temperature (300 K) and ambient density (1.8 g/cm$^3$). NVT simulations were performed for 200 ps, and the radial distribution functions of the molecules ($g_{r,mol}$) over the last 25 ps were computed. The resulting $g_{r,mol}$ along with snapshots are depicted in Fig. 4a and Fig. 4d. From both atomistic and CG simulations, we observe three peaks, P1, P2, P3 at ~4.3, 7.0, and 8.9

Å, respectively. These results demonstrate that the vast majority of molecules maintain the correct spatial group (*Pbca*) and lattice structures observed from the atomistic simulations, with a few molecules disordered. These results are significant because a previous CG force field for RDX using spline functions[30] reported that a 4-site model could not maintain the crystalline configurations above 30 K. Furthermore, with the exception of the recently published RDX-TCDD potential,[25] many of the CG force fields using 1-site models[17,30] led to hexagonally close-packed crystal structures instead of the *Pbca* space group, which contributes to the discrepancy in plasticity between those CG models and the atomistic model.[49]

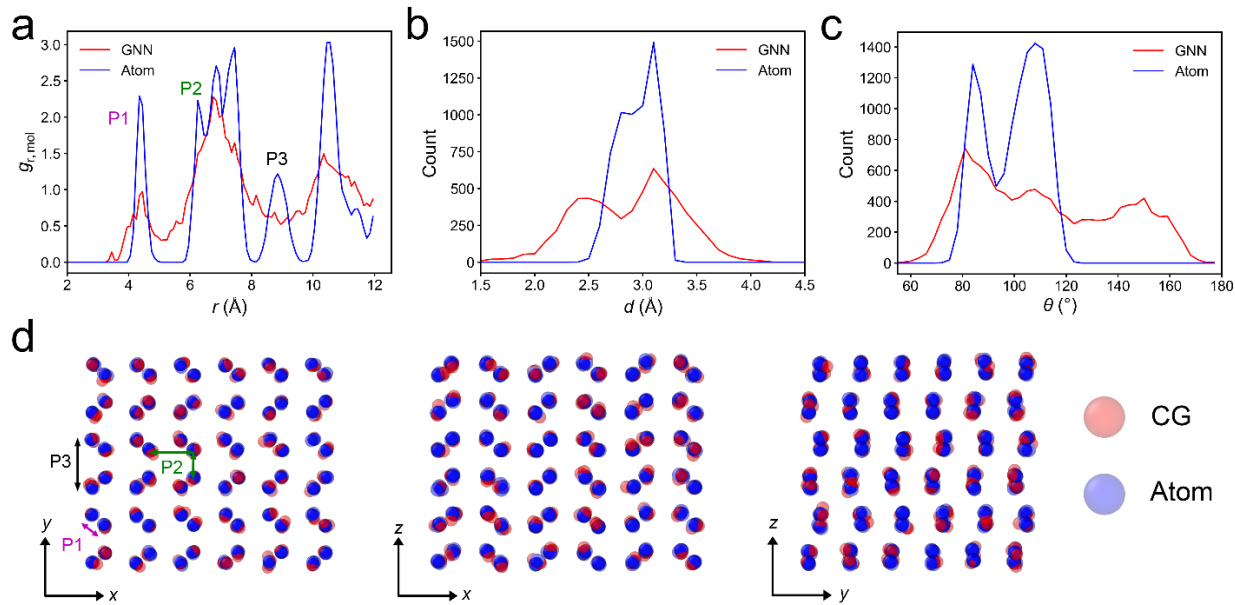

**Figure 4**. NVT simulation results for initially crystalline structures at 300 K and 1.8 g/cm$^3$. (a) $g_{r,mol}$, (b) intramolecular distances between nitro-triazine pairs, (c) intramolecular angles between nitro-triazine-nitro triplets, and (d) overlaying configuration snapshots of the molecules for the atomistic and CG-GNNFF models. The three nearest neighbor peaks from $g_{r,mol}$ (a) are depicted in the snapshot from the [001] perspective.

We further analyze the CG-GNNFF model by comparing to the intramolecular bond distances and angular distributions of the atomistic model. Typically, the RDX conformation is characterized by the out-of-plane bending or wag angle between the N-N bond between the nitro and triazine groups, and the plane formed by the C-N-C atoms in the triazine group.[50–54] α-RDX molecules consist of two N-N bonds in axial configurations and one bond in an equatorial configuration.[54] From atomistic trajectories of α-RDX that were converted to CG model center-of-mass representations, we observe intramolecular distances between nitro-triazine pairs in the range of 2.7 to 3.3 Å (Fig. 4b) with a sharp peak at 3.2 Å. In comparison, the CG-GNNFF models lead to broader distributions with peaks observed at 2.5 and 3.3 Å. For the intramolecular angles between the nitro-triazine-nitro triplets, we observe two distinct peaks at 110° and 90° from the atomistic simulations. The CG simulations display a similar peak at 90°, but a broader distribution above 110°.

Several factors may contribute to the broader distribution of the intramolecular configurations for the CG-GNNFF model. First, the CG-GNNFF model is trained on the mean force from atomistic simulations for a given CG configuration. Studies have reported that such mean-force models may underrepresent the frictional forces between molecules because the high-frequency motions of the hydrogen atoms are not explicitly captured.[41] Second, the stretching between the nitro and triazine group in the atomistic model represents two distinct molecular mechanics: the stiff stretching of the N-N bond whose stretching constant in the atomistic force field is 991.7 kcal/mol/Å, and the displacement of atoms within the groups that adjust their centers-of-masses. As the CG representation does not directly distinguish these two mechanisms, the intramolecular stretching with the CG force field may be softer than the instantaneous stretching of the N-N bonds in the atomistic model. Despite the broader distribution, we find that the CG-GNNFF performs well in describing the structure of RDX for a wide range of conditions as illustrated below.

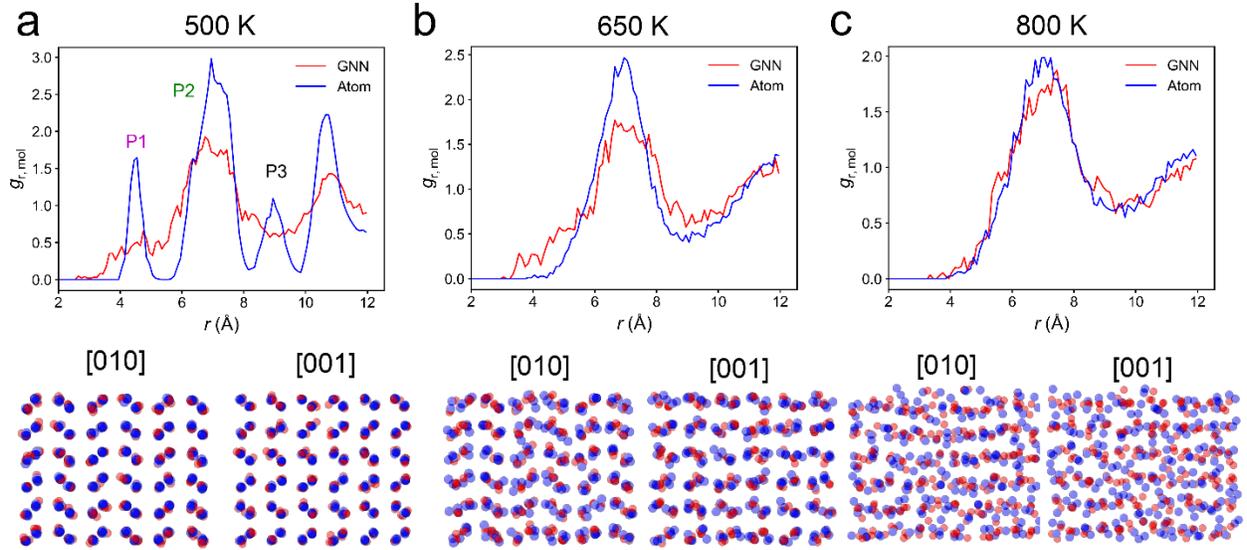

**Figure 5.** $g_{r,mol}$ and simulation configuration snapshots at ambient density and varying temperatures (a) 500 K, (b) 650 K, and (c) 800 K for 200 ps simulation times. In the snapshots below the plots in the perspective of the [010] and [001] orientations, red circles indicate the CG particles, while blue circles represent the atomistic model center-of-mass.

Next, we evaluate the temperature-dependent structural properties up to 800 K; we will contrast not just the equilibrium structures, but also the kinetics of the process of melting. From the atomistic simulations, we observe that the crystallinity is maintained over the 200 ps for the NVT simulations at temperatures up to 500 K. At 650 K, symmetry breaking and loss of crystalline order becomes evident by the disappearance of the P1 and P3 peaks in $g_{r,mol}$, as well as from snapshots viewed from [010] and [001] orientations. The CG model displays similar temperature dependence, where at 500 K, we observe that most of the CG particles remain at their lattice positions, although the P1 peak broadens as a few CG particles become slightly disordered. At 800 K, the P1 peak disappears completely resulting in a melt structure with a pair correlation function $g_{r,mol}$ in excellent agreement with the atomistic model. At the intermediate temperature of 650 K, the CG particles become partially disordered, but the P1 peak does not fully disappear within 200 ps, whereas in the atomistic simulations the peak disappears at ~30 ps (time-dependence of the atomistic model is not plotted).

We evaluate the extrapolation capability of the CG-GNNFF model to conditions outside the training dataset and study the structures at high density of 2.16 g/cm$^3$, prepared from NPT atomistic simulations with pressure of 5 GPa (Fig. 6a). The results demonstrate that the crystal structures of both the atomistic and CG models remain in the *Pbca* space group. The $g_{r,mol}$ peaks decrease from 4.3 Å and 7.0 Å for P1 and P2 at ambient (1.8 g/cm$^3$) density to 4.15 Å and 6.45 Å at 2.16 g/cm$^3$. Such changes are observed from both atomistic and CG simulations, demonstrating that CG-GNNFF can correctly capture density effects.

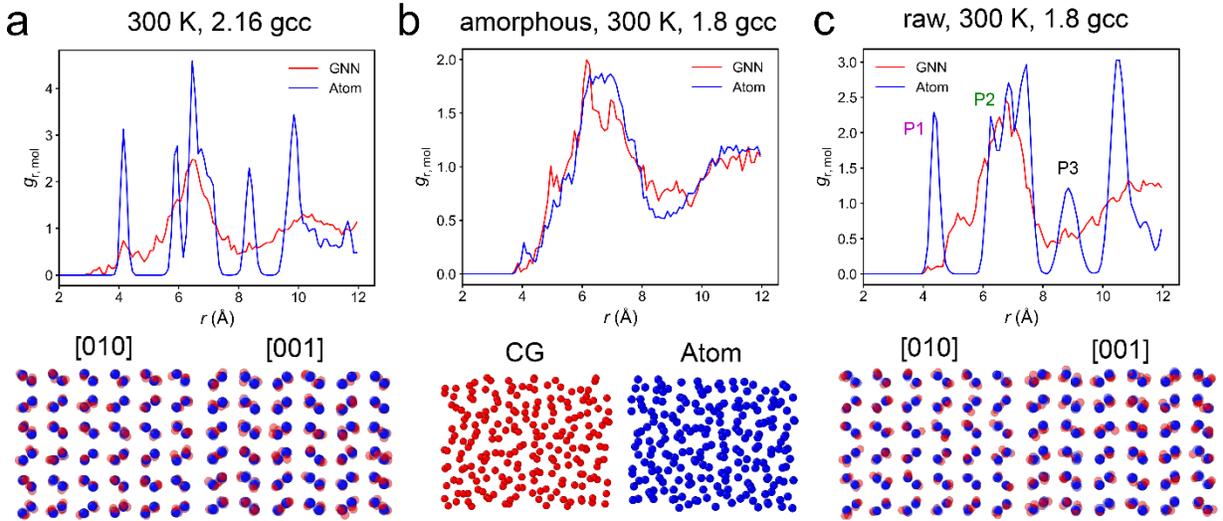

**Figure 6**. (a) $g_{r,mol}$ of a crystalline structure at density of 2.16 g/cm$^3$. (b) $g_{r,mol}$ of an initially amorphous sample. (c) $g_{r,mol}$ obtained from the CG model with raw coordinate input at 300 K and 1.8 g/cm$^3$. For the configuration snapshots shown below the plots for the [010] and [001] orientations, red circles indicate the CG particles, while blue circles represent the atomistic model centers-of-mass.

We also simulated the CG and atomistic models for configurations that were initially amorphous (Fig. 6b). We performed simulations for a range of temperatures (250-750 K) and densities (1.8-2.6 g/cm$^3$), and observe that the $g_{r,mol}$ functions are similar to those obtained from the structures at high temperatures (Fig. 5c) for simulations of initially crystalline configurations, marked by the disappearance of P1 peaks. The CG model captures this structural characteristic of amorphous configurations, as depicted in Fig. 6b.

Finally, we note that the molecular crystal structural properties are highly sensitive to the model formulation. For example, we have also trained a GNNFF model with raw coordinates of the interparticle distances and angles as inputs instead of the descriptors obtained from the Gaussian basis functions. Detailed descriptions of these models are given in SM section 1. The structural properties of the raw GNN model for initially crystalline, room temperature, and ambient density conditions are depicted in Fig. 6c. While the snapshots demonstrate that the majority of molecules are in their lattice positions in the *Pbca* space group, we see that the P1 peak is less distinct compared to the model with Gaussian basis functions, indicating more disordered molecules compared to results shown in Fig. 4a. This is because the intricate atomistic interactions that determine the structure of molecular crystals are not monotonous functions of the molecular configurations. RDX molecules display polymorphic phases because various interactions including stretching, bending, dihedral, improper dihedral, van der Waals, and coulombic forces of the diverse atoms act with distinct functional forms with respect to the molecular configurations. While in principle GNN can learn to describe these behaviors correctly with raw inputs, we find that guiding the model with Gaussian basis functions designed from the dataset is more beneficial in practice. Such findings demonstrate the intricacy of correctly describing the structure of molecular crystals.

*2.5 Model performance for faceted nanoparticles*

As an additional evaluation of the model performance for conditions outside its training dataset, we simulate the structural properties of faceted nanoparticles (NP) of RDX at varying temperatures. Such simulations require the model to capture free surfaces with asymmetric interaction geometries compared to the bulk systems that the model was trained upon. The faceted NPs were prepared according to the procedure outlined by Li et al.,[55] with the longest and shortest dimension lengths of 6 nm and 4 nm, respectively. We analyzed the structure of the molecules depending upon their positions and time for temperatures ranging from 300 to 600 K. To analyze surface phenomena, molecules were categorized as belonging to the core if they were within 2 nm of the center-of-mass of the NP, while belonging to the surface otherwise. Importantly, we analyzed not only the equilibrium state at various temperatures, but also the kinetics associated with the melting of the NPs.

Figure 7 compares radial distribution functions for the core and surface molecules and the molecular structures between the atomistic and CG models. As expected, the initially crystalline NPs lose crystallinity as the temperature is increased, where this process initiates at the free surfaces. We observe minor discrepancies between the atomistic and CG models in the core, where molecules of the atomistic model become disordered faster, while the surface molecules remain slightly more crystalline. Specifically, the core remains crystalline at 300 K and 400 K (Fig. 8a and 8b) for both the CG and atomistic models, with some disorder for the CG model demonstrated by a reduction in sharpness of the P1 peak at 200 ps. At 500 K, the core molecules of the atomistic simulations are partially disordered after 100 ps and completely amorphous after 200 ps (Fig. 8c), as well as at 600 K (Fig. 8d). The CG particles are partially disordered throughout the simulation at 500 K, and up to 100 ps for the 600 K simulations. At 600 K and 200 ps, the core CG particles are completely amorphous, matching the behavior observed from the atomistic simulations. For the surface molecules, the atomistic model leads to a crystalline structure for the duration of the simulation at 300 K, whereas the CG model leads to partially disordered structures at 200 ps (Fig. 8a). Above this temperature, the surface molecules are mostly in amorphous states for both models. Overall, similar to the bulk simulation behavior, we observe that the CG model simulations for the faceted NPs lead to similar but more disordered structures compared to the atomistic model simulations.

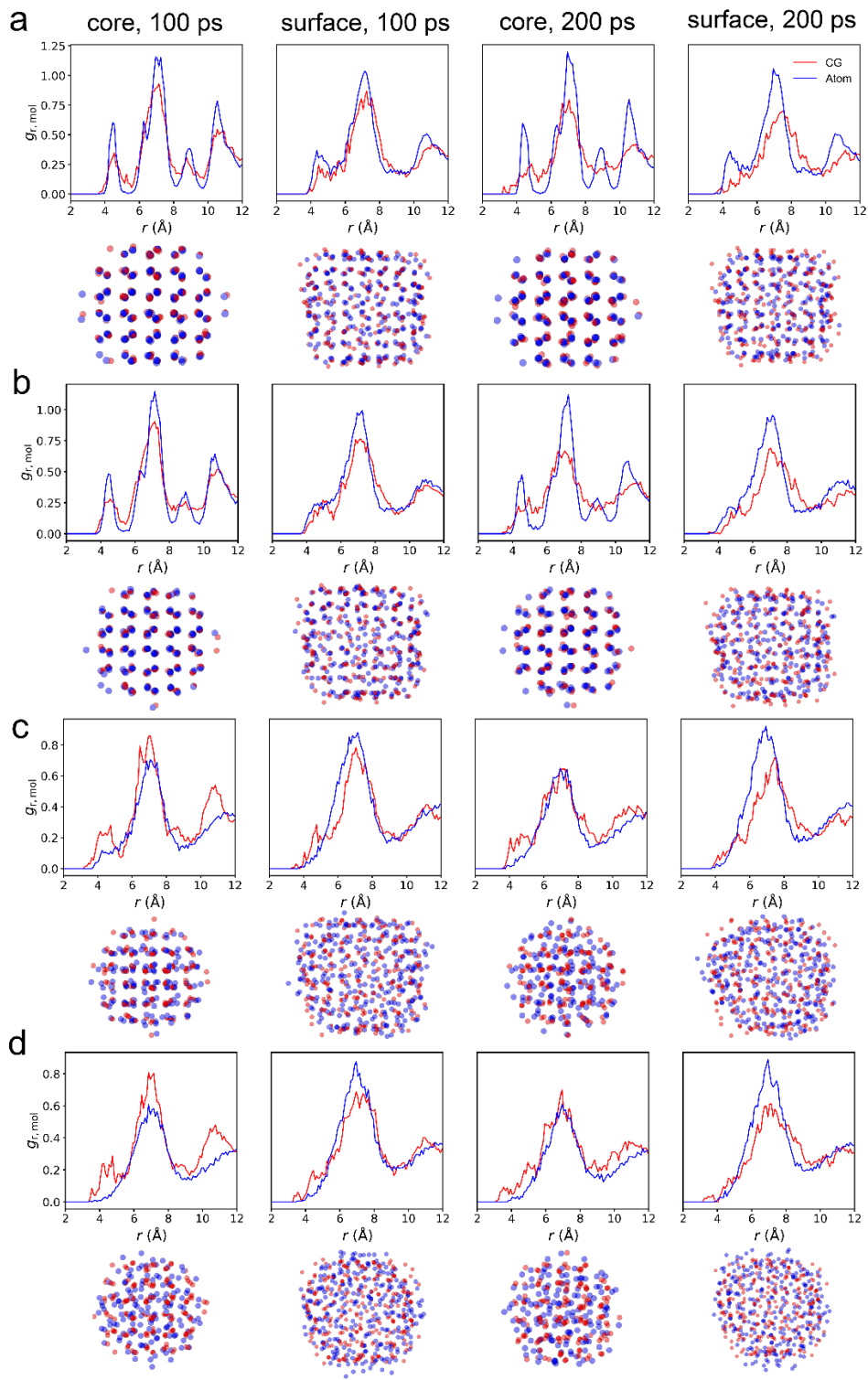

**Figure 7**. $g_{r,mol}$ of faceted nanoparticles of RDX at $T$ of (a) 300 K, (b) 400 K, (c) 500 K, and (d) 600 K. The core molecules are within 2 nm from the center-of-mass, while those outside this range are considered surface molecules. The snapshots are for the perspective of the [010] orientation.

## 3. Discussion

In this study, we present a CG force field for a molecular crystal that can capture both low-symmetry crystal structures as well as amorphous structures. The graph neural network force field was formulated to conserve the total linear and angular momentum from the intramolecular interactions that include two- and three-body interactions, which has not been previously considered in direct force prediction models.[38,56] Gaussian basis functions were designed based on the molecular configurations to emphasize the crystal symmetry. The CG model was iteratively improved by expanding the training dataset through the development of a neural network decoder that reconstructs atomistic model configurations from CG model trajectories.

While the GNN model captures both the crystal and amorphous structures at a wide range of densities and temperatures including the structure of nanoparticles that were not in the training dataset, it also displays discrepancies in several aspects. For example, we see that the radial distribution peaks for the crystalline configurations are not as sharp for the CG model because some particles become disordered. In addition, there is a significant difference in the dynamics of the two models as demonstrated by the mean squared displacement over time depicted in Fig. S5 in SI. Such deviation is a well-documented phenomena in previous studies of CG models[22,41,57–60] that could be alleviated through the use of CG model methodologies such as incorporating dissipative particle dynamics or scaling of dynamics as noted by Brennan et al.[61] Such differences in dynamics may be responsible for the discrepancies in structures as our observation time is fixed at 200 ps. For example, while the reported melting temperature of RDX is 488 K,[62] our atomistic 200-ps simulations at 500 K show a rigidly crystalline structure indicating that the structure has not reached its thermodynamically stable configuration in the simulated time frame. Therefore, the discrepancy in structural properties between the atomistic and CG models may be due to the difference in the kinetics of the two models. Overall, we attribute these deficiencies to the design of the CG model. Even for the ideally trained model, the CG model describes the mean force of a group of atoms at ps time scales, whereas the time steps for the CG simulations are in the fs scales. With such data curation protocol, the mean forces are obtained from energy minimized configurations, leading to underrepresentation of energetically unfavorable configurations that may act as energetic barriers against certain configurational transitions. Therefore, the resolution and time-scale of the CG model prevents a precise description of key factors such as frictional forces or details of the rugged free-energy landscape of the molecular crystals.

## 4. Methods

### 4.1 Graph transformer neural network

The graph transformer neural network[43] is used for the CG-GNNFF. In this model, the query ($q_{c,i}^l$), key ($k_{c,j}^l$), value ($v_{c,j}^l$), and edge ($e_{c,ij}^l$) vectors for the edge between node $i$ and $j$ of GNN layer $l$ is given by,

$$q_{c,i}^l = W_{c,q}^l h_i^l + b_{c,q}^l \tag{2}$$

$$k_{c,j}^l = W_{c,k}^l h_j^l + b_{c,k}^l \tag{3}$$

$$v_{c,j}^l = W_{c,v}^l h_j^l + b_{c,v}^l \tag{4}$$

$$e_{c,ij}^l = W_{c,e}^l e_{ij} + b_{c,e}^l \tag{5}$$

Here, $q, k, v$ are vectors with dimension $H$ equal to the size of the hidden embedding for the layer, while $W$ and $b$ are trainable parameters. $c$ represents the index of the attention head. The multi-headed attention ($\alpha^l_{c,ij}$) is then calculated as,

$$\alpha^l_{c,ij} = \frac{\langle q^l_{c,i}, k^l_{c,j} + e^l_{c,ij} \rangle}{\sum_{u \in \mathcal{N}(i)} \langle q^l_{c,i}, k^l_{c,u} + e^l_{c,iu} \rangle} \tag{6}$$

where $\langle q, k \rangle = \exp\left(\frac{q^T k}{\sqrt{d}}\right)$.

Messages for node $i$ are aggregated from all its neighbors ($\mathcal{N}$) according to,

$$h^l_i = \tanh\left[\sum_{c=1}^{C} \sum_{j \in \mathcal{N}(i)} \alpha^l_{c,ij}(v^l_{c,j} + e^l_{c,ij})\right], \quad l \neq L \tag{7}$$

The message passing procedure for the last layer ($l = L$) differs from the other layers as it incorporates a force direction vector ($u_{ij}$, described in section 2.1) that specifies the node features partitioning into the force directions. Therefore, the force on each CG bead ($F^{GNN}$) is determined in the last layer according to

$$F^{GNN}_{i,k} = \sum_{c=1}^{C} \sum_{j \in \mathcal{N}(i)} \alpha^l_{c,ij}(v^l_{c,j} + e^l_{c,ij}) * u_{ij,k}, \quad l = L \tag{8}$$

where the index $k \in (x, y, z)$ indicates the Cartesian axis of the force. For intermolecular forces, Newton's third law of motion is enforced by formulating the forces as $F^{GNN}_{i,k} = \sum_{c=1}^{C} \sum_{j \in \mathcal{N}(i)} [\alpha^l_{c,ij}(v^l_{c,j} + e^l_{c,ij}) + \alpha^l_{c,ji}(v^l_{c,i} + e^l_{c,ji})]/2 * u_{ij,k}$.

The hyperparameters of the GNN are the size of the hidden embeddings ($H$), number of attention heads ($C$), and number of message passing layers ($L$). We chose $H=8$, $C=5$, and $L=4$ because such hyperparameters led to good accuracy in force prediction, while increasing the number of parameters did not lead to a significant increase in the prediction accuracy.

The GNNs were trained with a stochastic gradient descent optimizer with momentum of 0.7 and learning rate of 0.005 for 2000 epochs. The loss function was the mean squared error between the GNN output ($F_x^{GNN}, F_y^{GNN}, F_z^{GNN}$) and the CG model forces obtained from the atomistic model simulations ($F_x^{CG}, F_y^{CG}, F_z^{CG}$). The training and test dataset were split 80/20.

*4.2 Atomistic and CG molecular dynamics*

Atomistic model simulations of RDX utilized the Smith and Bharadwaj (SB) potential[63] that has been shown to accurately capture the structural properties of RDX. Crystalline configurations of RDX were prepared by creating a 3 × 3 × 3 supercell of α-RDX, whose initial configuration was obtained from the Cambridge Crystallographic Data Centre.[64] Samples were equilibrated by running isobaric and isothermal (NPT) ensemble simulations for 10 ps, followed by production simulations conducted within the canonical ensemble (NVT). The temperatures ranged from 250 K to 800 K and pressures ranged from 0 to 20 GPa. The Nosé-Hoover thermostat and barostat were used. All atomistic simulations were performed with Large-scale Atomic/Molecular Massively Parallel Simulator (LAMMPS).[65]

In addition to the crystalline configurations, amorphous structures were prepared following the procedure by Sakano et al.[66] First, crystalline samples were heated to 800 K for 200 ps and equilibrated at 800 K for 300 ps at ambient pressure using the NPT ensemble. Next, the simulation cells were deformed under the NVT ensemble to match the densities obtained from NPT simulations of crystalline RDX.

For the CG model simulations, we obtain initial configurations from the center-of-masses of the nitro and triazine groups from the atomistic model trajectories. Langevin dynamics (LD) were utilized to run the NVT simulations. CG simulation codes were written in-lab with Python. The Pytorch Geometric package with necessary modifications for our model was used to build the graph neural network. The code for the CG simulations is provided in github.[67]

## Data availability

Data used to train and test the CG-GNNFF as well as the best model are available along with the source code at https://github.itap.purdue.edu/StrachanGroup/CG_GNNFF. Additional data is available from the corresponding authors on reasonable request.

## Code availability

The codes associated with the model training, decoder to reconstruct CG configurations from atomistic trajectories, and Langevin dynamics are available at:

https://github.itap.purdue.edu/StrachanGroup/CG_GNNFF.

## Supplementary material

The supplementary material includes detailed explanation of the model input features and associated weights, description of the training dataset, employed prior forces, generation dependent force prediction accuracy, and mean squared displacement data.

## Acknowledgments

The authors thank Brenden Hamilton for many insightful discussions and comments. This research was sponsored by the Army Research Laboratory and was accomplished under Cooperative Agreement Number W911NF-20-2-0189. This work was supported in part by high-performance computer time and resources from the DoD High Performance Computing Modernization Program. The views and conclusions contained in this document are those of the authors and should not be interpreted as representing the official policies, either expressed or implied, of the Army Research Laboratory, or the U.S. Government. The U.S. Government is authorized to reproduce and distribute reprints for Government purposes notwithstanding any copyright notation herein.## Author declarations

The authors have no conflicts to disclose.

# Supplementary Material: Graph neural network force field for coarse-grain molecular crystal

Brian H. Lee,[1] James P. Larentzos,[2] John K. Brennan,[2] Alejandro Strachan[1, a)]

Author affiliations
[1]*School of Materials Engineering and Birck Nanotechnology Center, Purdue University, West Lafayette, Indiana 47907, USA*
[2]*U.S. Army Combat Capabilities Development Command (DEVCOM) Army Research Laboratory, Aberdeen Proving Ground, Maryland 21005, USA*

Author email
[a)]Author to whom correspondence should be addressed: strachan@purdue.edu

## 1. Model features

In training the GNNFF, we have experimented with two types of edge features and two types of node features. For the first node feature type, the intramolecular node features were two-dimensional one-hot vectors indicating whether the particle is nitro or triazine group. For the second node feature type, the node features had an additional dimension that recorded the local density of the particle, whose definition is given by Moore et al.[1],

$$\rho_{local,i}(r_{ij}) = \frac{84}{5\pi r_{cut}^3}\left(1 + \frac{3r_{ij}}{2r_{cut}}\right)\left(1 - \frac{r_{ij}}{r_{cut}}\right)^4 \tag{1}$$

Here, $\rho_{local,i}$, $r_{ij}$, $r_{cut}$ are the local density of particle $i$, distance between particles $i$ and $j$, and the cutoff distance corresponding to 9 Å in our model, respectively. We find that the density feature does not significantly affect the crystal and amorphous structures observed from the CG simulations. Therefore, we use this feature only for data generation as explained in section 2 of Supplementary Material (SM). All of the results in the main manuscripts are from the first node feature type without the local density as input.

The types of edge features were varied to either accept raw coordinates or an expansion of the coordinates in Gaussian bases. The edge features of intramolecular stretching interactions ($e_{str}$) with raw coordinates were three-dimensional vectors equal to $[d_{N_i-T}, \theta_{N_i-T-N_j}, \theta_{N_i-T-N_k}]$. The intramolecular angular features ($e_{ang}$) were $[\theta_{N_i-T-N_j}, d_{N_i-T}, d_{T-N_j}]$ and the intermolecular features ($e_{inter}$) were $[f_c(d)]$, where $f_c$ is a smoothly decaying function defined by Behler et al.[2],

$$f_c(d_{ij}) = \begin{cases} 0.5[\cos(\pi d_{ij}/r_{cut}) + 1] & \text{for } d_{ij} \leq r_{cut} \\ 0 & \text{for } d_{ij} > r_{cut} \end{cases} \tag{2}$$

Here, $r_{\text{cut}}$ is the cutoff distance and is defined as 9 Å in this study to include the first three peaks of the radial distribution function of crystalline RDX.

For the edge features with Gaussian bases, we expand the raw features with four Gaussian functions. For example, the intermolecular edge features were $[G_{1,\text{inter}}(f_c(d)), G_{2,\text{inter}}(f_c(d)), G_{3,\text{inter}}(f_c(d)), G_{4,\text{inter}}(f_c(d))]$. The intramolecular edge features were 12-dimensional vectors that expand the raw features with Gaussian functions. The Gaussian functions are defined as,

$$G_n(x) = e^{-\eta_n (x - x_{0,n})^2} \tag{3}$$

The parameters for the Gaussian functions $(\eta_n, x_{0,n})$ are given in Table S1. These values are chosen based on the inter- and intramolecular configurations in our training dataset described in section S2, Fig. S1.

**Table S1. Gaussian function parameters**

| Type | $\eta$ | $x_0$ |
| --- | --- | --- |
| $G_{1,\text{str}}$ | 5 | 2 Å |
| $G_{2,\text{str}}$ | 20 | 2.7 Å |
| $G_{3,\text{str}}$ | 20 | 3.1 Å |
| $G_{4,\text{str}}$ | 5 | 4 Å |
| $G_{1,\text{ang}}$ | 10 | 1.047 rad |
| $G_{2,\text{ang}}$ | 20 | 1.395 rad |
| $G_{3,\text{ang}}$ | 5 | 2.094 rad |
| $G_{4,\text{ang}}$ | 10 | 2.793 rad |
| $G_{1,\text{inter}}$ | 0.15 | 0 Å |
| $G_{2,\text{inter}}$ | 2 | 3.5 Å |
| $G_{3,\text{inter}}$ | 5 | 5.5 Å |
| $G_{4,\text{inter}}$ | 5 | 6.4 Å |

## 2. Training dataset

The training dataset is obtained from the iterative training approach explained in Section 2.2 of the main manuscript. We use three generations of data. Gen 0 is obtained from atomistic model simulations of crystalline RDX at ambient density (1.8 g/cm$^3$) and temperatures of 250 K, 500 K, and 750 K. Gen 1 and Gen 2 datasets are obtained from CG model simulations with the same conditions as Gen 0. Gen 1 includes four sets of trajectories obtained from four models with varying edge and intramolecular node features outlined in Section S1. The Gen 2 dataset is obtained from two models that both utilize Gaussian functions for the input features, while varying the intramolecular node features. In total, Gen 0, Gen 1, and Gen 2 contained 300, 1073, and 540 frames, respectively, for a total of 1.6 million data points. We also note that some of the configurations obtained from the CG model simulations were not included in the dataset because the decoded atomistic model trajectory led to numerically unstable configurations.

Fig. S1 depicts an analysis of the configuration distributions within the training datasets. Here, the intramolecular configuration distributions are counts of the configurations normalized by the maximum number of counts. The intermolecular configuration distributions are counts of the intermolecular distances that are multiplied by the smoothing function ($f_c$) and normalized by the maximum count. We observe that the following configurations are under-represented in the dataset: intramolecular stretching configurations of ( $d > 3.6$ Å and $d < 2.0$ Å ), angular configurations of ($\theta < 60°$, $\theta > 160°$), and intermolecular overlapping configurations of ($d <$ 2.5 Å). Therefore, we utilize prior forces to avoid exploration of such configurations in the CG model simulations.

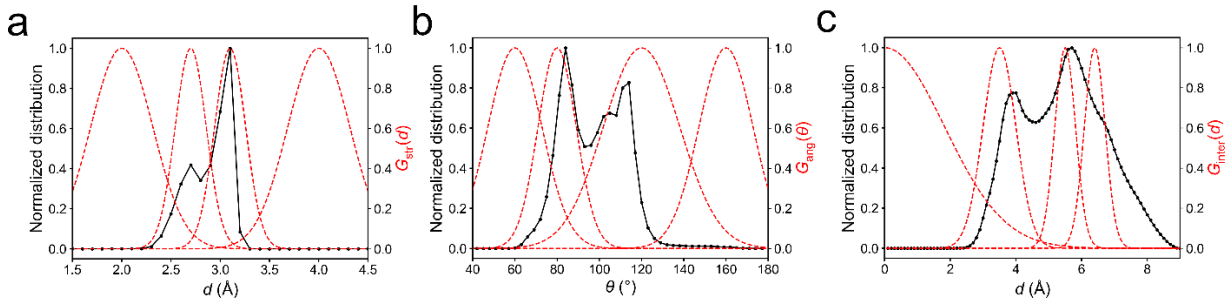

**Figure 1**. Normalized distribution of configurations in the training dataset and the associated Gaussian functions. (a) Intramolecular distances, (b) intramolecular angles, (c) intermolecular distances. Black lines are the normalized counts of configurations. Red dashed lines are the Gaussian functions.

## 3. Prior forces

Simulations employing neural network force fields are prone to exploring configurational spaces that are significantly under-represented in the training data set. For these configurations, the extrapolation of a GNNFF is not guaranteed to be accurate and can lead to catastrophic failures of the model. To avoid these cases, we employ prior forces that restrict the exploration of such configurations following the approach by Wang et al.[3] Specifically, we apply prior forces for the intramolecular stretching configurations ($d > 3.6$ Å and $d < 2.0$ Å), angular configurations ($\theta < 60°$, $\theta > 160°$), and intermolecular overlapping configurations ($d < 2.5$ Å), as these configurations are not sampled sufficiently in the training set (see Fig. S1). The prior forces for these configurations are determined from the potential of mean force (PMF) calculations obtained from steered molecular dynamics of the atomistic model, gas phase RDX using the Colvar package[4] of LAMMPS.

For the intramolecular interaction prior forces, the nitro-triazine distance (Fig. 2a) and the nitro-triazine-nitro angle (Fig. 2b) of a single RDX molecule were used as the collective variables. For the intermolecular interactions (Fig. 2c), the distance between the nitro groups of two RDX molecules were used as the collective variable. In all cases, the PMF in the configurations of interest for prior forces can be described by harmonic potentials, $U = \frac{k}{2}(r - r_0)^2$. Therefore, we determine the prior forces as given by the following equations,

$$F_{\text{str}} = \begin{cases} 159.08(d - 3.6) & \text{for } d > 3.6 \text{ Å} \\ 176.09(d - 2.0) & \text{for } d < 2.0 \text{ Å} \end{cases} \quad (4)$$

$$F_{\text{ang}} = \begin{cases} 0.0812(\theta - 60) & \text{for } \theta < 60° \\ 0.0151(\theta - 160) & \text{for } \theta > 160° \end{cases} \quad (5)$$

$$F_{\text{inter}} = 105.22(d - 2.5) \quad \text{for } d < 2.5 \text{ Å} \quad (6)$$

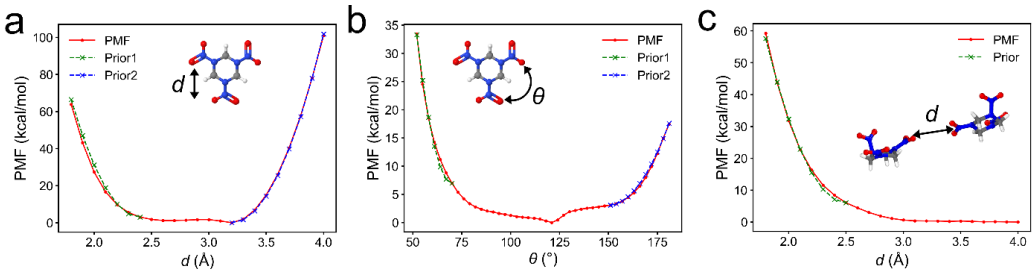

**Figure 2**. PMF from the atomistic model and corresponding prior energies for (a) intramolecular stretching, (b) intramolecular bending, and (c) intermolecular overlapping configurations.

## 4. Generation-dependent inference

Generation-dependent inference performance of the intra- (Fig. 3) and inter-molecular (Fig. 4) forces are depicted. Here, the rows correspond to the generation dataset that the model was trained on. For example, second row plots correspond to inference of the model that was trained with all data from Gen 0 and Gen 1. The columns correspond to the generation dataset that the model was used for inference. For example, the middle column plots correspond to inference on Gen 0 and Gen 1 datasets with model trained on Gen 0 data (first row) and Gen 0 and Gen 1 data (second row). Only test set results are plotted as the model performs similarly for the train set.

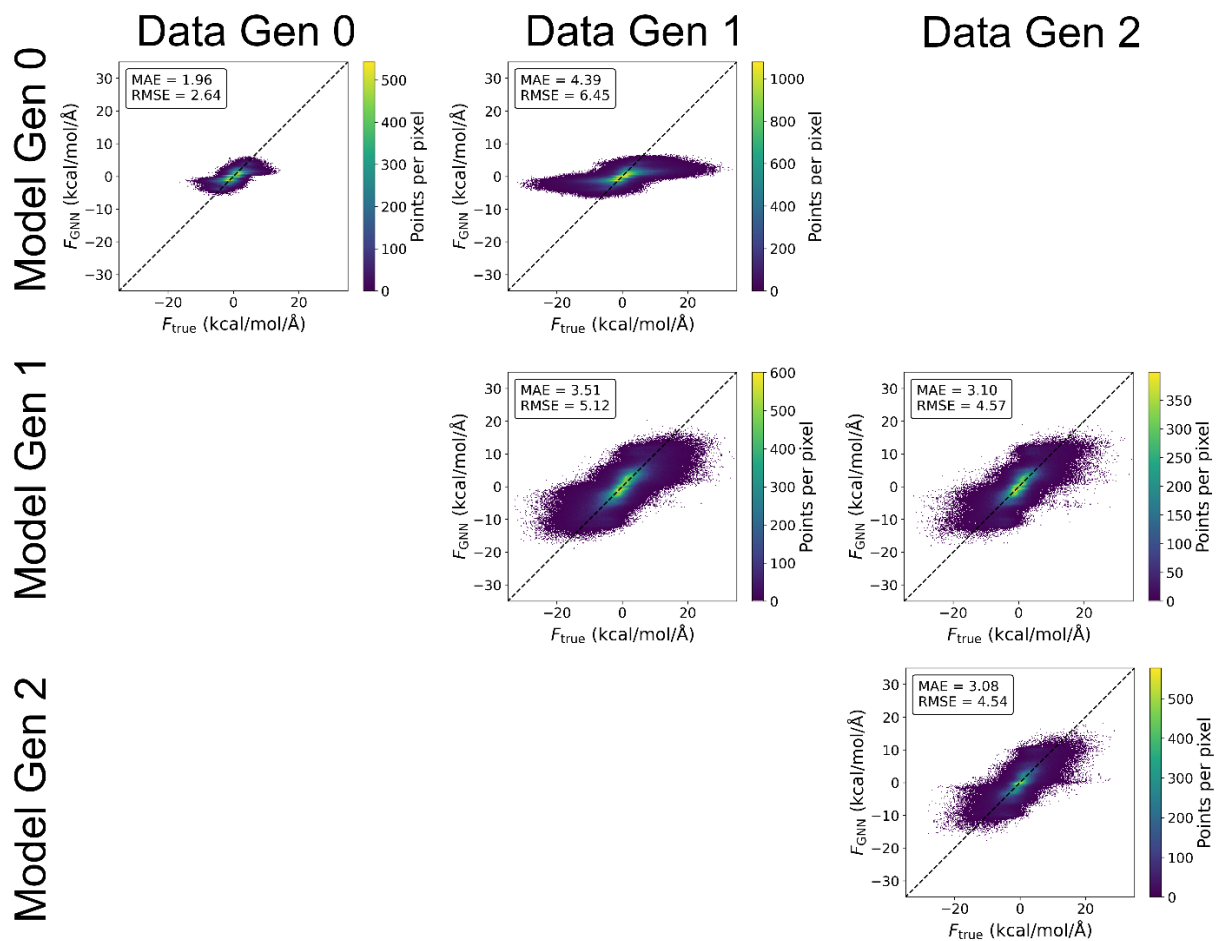

**Figure 3**. Generation dependent inference performance for intramolecular forces. The columns correspond to the dataset generation used for inference, while the rows correspond to the dataset used for training the model.

The results demonstrate that the Gen 0 force magnitudes are smaller compared to later generations. This is because the molecular configurations are from the crystalline configuration of the atomistic model simulations that do not deviate significantly from the local energy minima configurations. However, as the generations are added, configurations with large forces are added.

The model trained on Gen 0 does not perform well on such configurations and predicts them to have small force magnitudes. However, the models trained with those iteratively added configurations are able to predict the large magnitude forces.

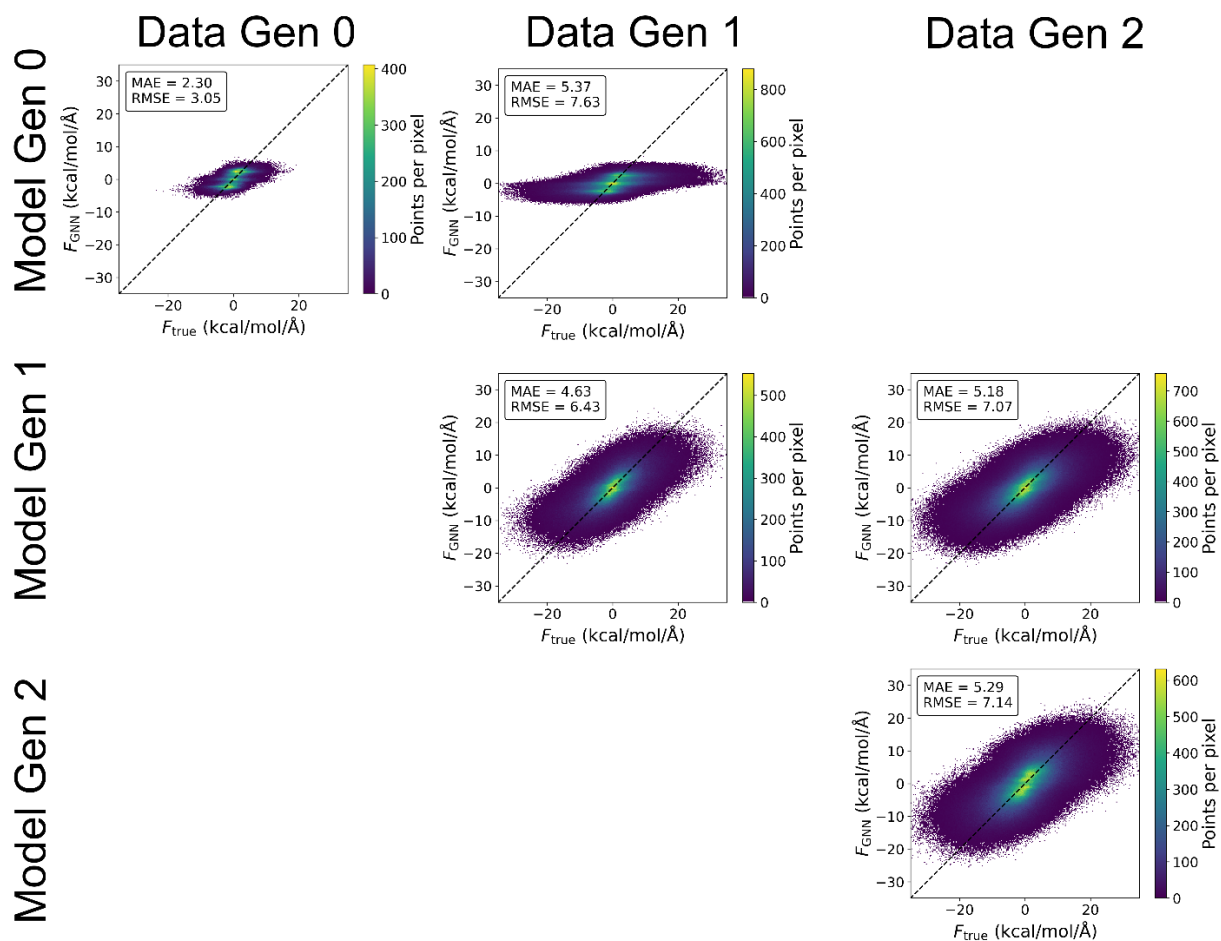

**Figure 4**. Generation dependent inference performance for intermolecular forces. The columns correspond to the dataset generation used for inference while the rows correspond to the dataset used for training the model.

## 5. Dynamic properties

The mean squared displacements (MSD) of the center-of-mass of molecules were measured for bulk system atomistic and CG model simulations. As reported for previous CG models,[5–8] there is significant difference in the dynamic properties that is not resolved by our CG model.

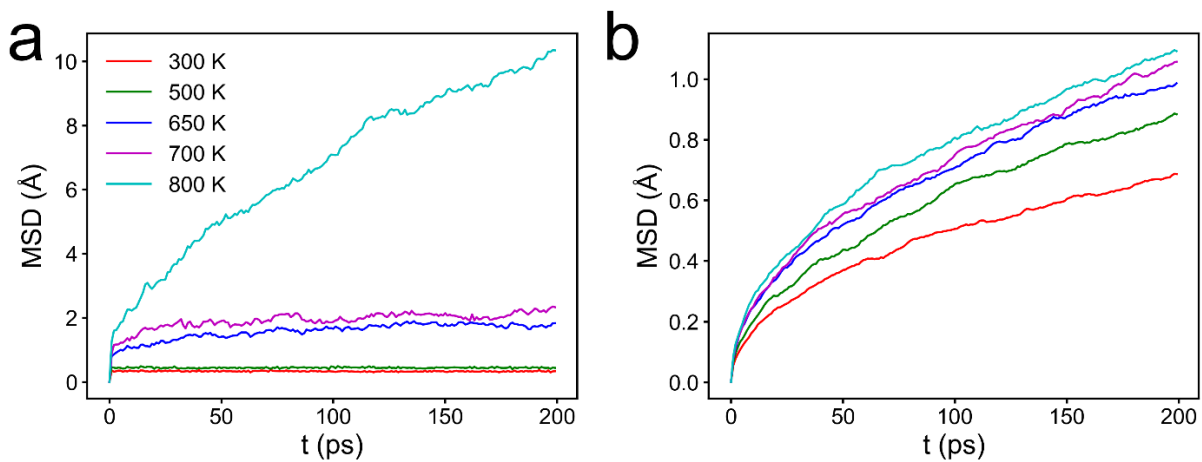

**Figure 5**. MSD from bulk simulations at ambient densities and varying temperatures for the (a) atomistic and (b) CG models.